# A Survey of Detection Methods for Die Attachment and Wire Bonding Defects in Integrated Circuit Manufacturing

Lamia Alam and Nasser Kehtarnavaz
Department of Electrical and Computer Engineering
University of Texas at Dallas
Richardson, TX 75080
lamia.alam@utdallas.edu, kehtar@utdallas.edu

*Abstract*— Defect detection plays a vital role in the manufacturing process of integrated circuits (ICs). Die attachment and wire bonding are two steps of the manufacturing process that determine the power and signal transmission quality and dependability in an IC. This paper presents a survey or literature review of the methods used for detecting these defects based on different sensing modalities used including optical, radiological, acoustical, and infrared thermography. A discussion of the detection methods used is provided in this survey. Both conventional and deep learning approaches for detecting die attachment and wire bonding defects are considered along with challenges and future research directions.

*Keywords— integrated circuit (IC) defects, die attachment and wire bonding defects, defect detection in IC manufacturing*

## I. Introduction

Integrated circuits (IC) are used in nearly all electronic products including smartphones, personal computers, medical devices, cars, etc. By shrinking the sizes of electronic circuit elements from micrometer to nanometer, the capacity of ICs has increased dramatically. IC manufacturing has become one of the fastest-expanding technologies due to the demand for smaller size electronic products. During their manufacturing process, defects in ICs impact production costs negatively [1]. Defect detection as part of the IC manufacturing quality control is done for the purpose of lowering production costs [2, 3].

There are two types of inspection methods or testing which are used for examining ICs: destructive and non-destructive [4]. Destructive testing (DT) involves those tests which break down the material of an IC for the purpose of determining its physical properties, such as mechanical attributes of hardness, strength, flexibility, and toughness. Decapsulation, scribing metal, and cross sectioning constitute destructive techniques noting that when they are performed, they result in irreversible damage. Normally destructive testing is conducted before an IC enters mass production to obtain its limits in order to maintain proper operating conditions for the IC manufacturing machines. On the other hand, non-destructive testing (NDT) allows the examination of ICs without causing any damage to their material. Non-destructive testing techniques have been integrated into the production line of ICs for monitoring their quality. NDT plays a crucial role in the IC manufacturing industry.

Die attachment and wire bonding are two important steps of the IC manufacturing process as they set the power and signal transmission quality in ICs. The focus of this paper is placed on these two steps. The sensing modalities that are used for detecting defects of die attachment and wire bonding includes optical, radiological, acoustical, and infrared thermography [4].

The objective of this paper is to provide a survey or review of papers that address the detection of faulty die attachment and wire bonding via different sensing modalities. In Section II, an overview of the IC manufacturing process is initially provided. Then, in Section III, the sensing modalities that have been used for detecting IC defects are stated. Section IV covers various approaches that are introduced in the literature for the detection of die attachment and wire bonding defects. Challenges and future research directions are then stated in Section V. Finally, the paper is concluded in Section VI.

## II. Overview of IC Manufacturing Process

The manufacturing process of ICs can be divided into the following three major steps: formation of silicon wafer, wafer fabrication, and assembly/testing. An overview of these steps is provided next in order to set the stage for the defect detection survey presented in this paper.

### A. Formation of Silicon Wafers

ICs are made on semiconductor wafers which are made up of silicon. Figure 1 depicts the steps involved in the formation of silicon wafers. For building a semiconductor wafer, a silicon ingot is first grown using the Czochralski (CZ) method [5], that is heating silicon above its melting point in an Argon atmosphere under vacuum. Donor impurity atoms then get added, called doping, in precise amounts. After a polycrystalline and dopant mixture has liquified, a single silicon crystal known as the seed is placed on top of the melt, just touching the surface. The seed crystal is taken out of the melt once it has achieved the correct stage for crystal formation. The quick pulling of the seed crystal starts the growth process. The pull speed is then lowered to allow the crystal diameter to grow. The seed is carefully lifted above the melt once it has attained the proper diameter and the growing parameters have been stabilized to keep it there. After the ingot has grown to its full size, it is grounded to a rough size diameter that is somewhat bigger than the final silicon wafer's target diameter. The ingot is then sliced after going through a number of examinations. After slicing, the wafer goes through a lapping process to eliminate saw marks and surface flaws from both the front and back sides. Then, it goes through etching (chemical polishing) and cleaning. An edge grinding procedure is carried out to round the edges in order to reduce the possibility of breakage in the remaining formation steps. Next, wafer polishing is done in a clean room to produce a mirror finish. Then, a cleaning process is carried out in two stages: stock removal and a final chemical mechanical polish. Polishing pads and polishing slurry are used in both stages. After completing the final cleaning stage, a silicon wafer is sorted using either a manual inspection under high-intensity lighting or a laser scan. Finally, silicon wafers that match the requirements are bundled in cassettes and taped.

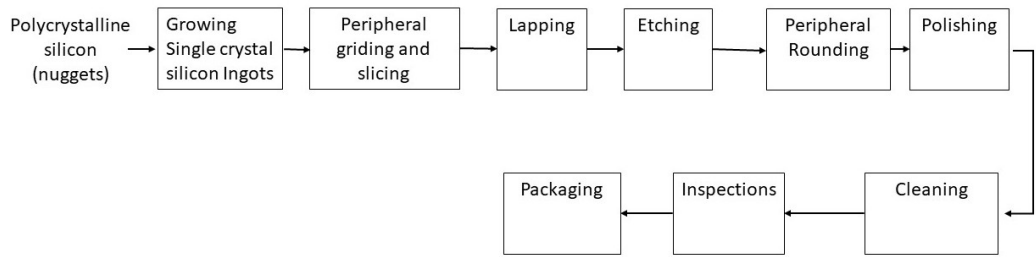

Fig. 1. Silicon wafer formation steps

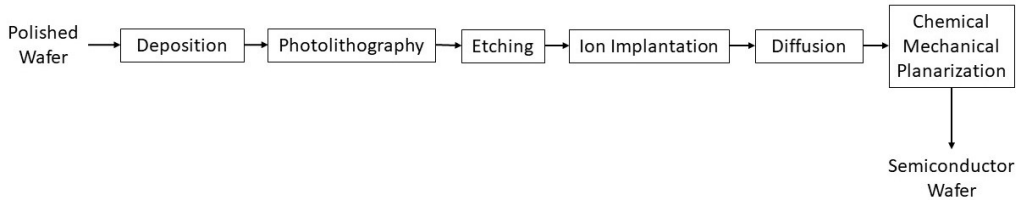

Fig. 2. Wafer fabrication process

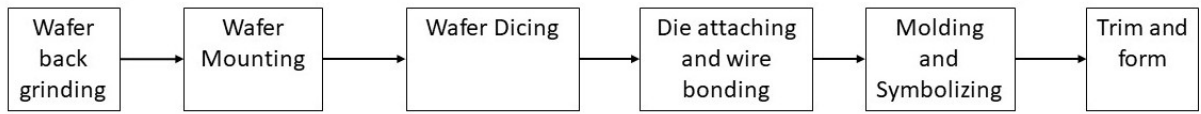

Fig. 3. Assembly and testing in IC manufacturing process

*B. Wafer Fabrication*

For fabrication, a silicon wafer goes through several steps which are depicted in Figure 2. Deposition and oxidation are the first steps. Layers of materials are deposited. When exposed to light, a circuit pattern on a mask is transferred via a photoresist layer, which changes its physical properties. Photolithography is the process of transferring a pattern from a photomask to the wafer's surface. The pattern gets imprinted on a photoresist layer on top of the wafer. Next, etching is carried out to remove materials in a selective manner towards creating a pattern defined by the etching mask to protect the parts that are to remain. Wet (chemical) or dry (physical) etching is then applied to remove the exposed material. The most common method for introducing dopant impurities into crystalline silicon is ion implantation. Diffusion is used to anneal crystal defects following ion implantation. Etching, deposition, and oxidation processes change the topography of the wafer surface, resulting in a non-planar surface. With the use of a chemical slurry, chemical mechanical planarization (CMP) is carried out to plane the wafer surface.

*C. Assembly and Testing*

To manufacture an IC, a die preparation step is done which typically begins with back grinding a semiconductor wafer. Wafer back grinding removes material from the backside of a wafer to a desired final target thickness. Wafer mounting is also done during the die preparation step. During this phase, a wafer is placed on a plastic tape that is attached to a ring. Wafer mounting takes place just before a wafer is sliced into individual dies. Wafer dicing involves separating dies from a semiconductor wafer. The dicing step involves scribing, breaking, and mechanical sawing or laser cutting. Die attachment is performed after the dicing step in order to fix an IC onto a substrate and its metal frame. Then, wire bonding is done to establish transmission of electrical signals. An IC chip is finally mounted on a ceramic or plastic package with logo and other information getting printed on it. Figure 3 depicts the assembly and testing steps of the IC manufacturing process.

III. SENSING MODALITIES

The IC manufacturing industry constantly takes steps to improve production efficiency and thus to reduce production costs. For testing and inspection of ICs, the sensing modalities of optical, radiological (x-ray), acoustical, and infrared thermography have been used.

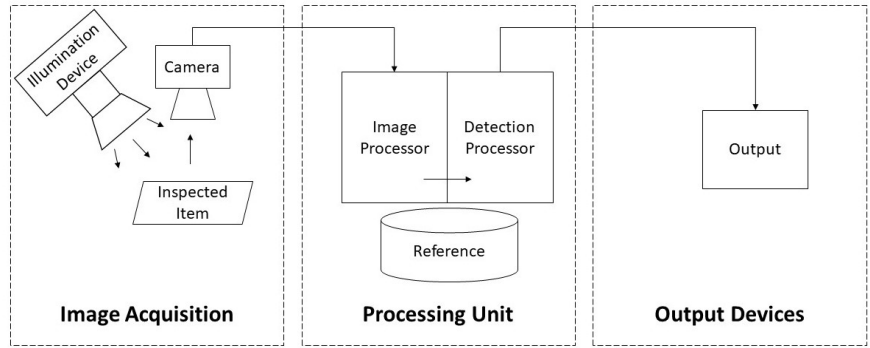

Fig. 4. Illustration of an AOI System

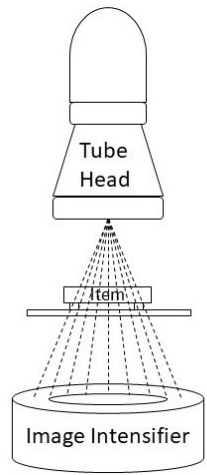

Fig. 5. Illustration of an x-ray inspection system

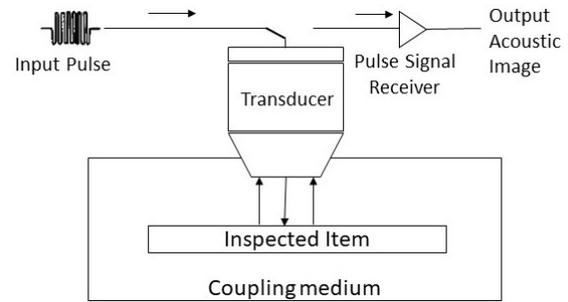

Fig. 6. Illustration of SAM inspection system

### A. Optical

Optical sensing for defect detection is a widely used sensing modality. Manual optical inspection is performed by a human inspector and automatic optical inspection (AOI) is performed by an automated system containing an image sensor such as a camera [6]. AOI is preferred over manual inspection due to the dull and fatigue aspects of manual inspection. Figure 4 illustrates a depiction of an AOI system.

A typical AOI system has three components: image acquisition, image processing, and output device. In such a system, light gets projected onto an IC by a special illumination source. The reflection of light from the IC is then collected by high quality lenses of a camera that generates an image of the IC. Motors are used to move the components of an AOI system for seeing all the parts of an IC. For defect detection, a computer program is used to analyze the captured image. The output of the system consists of a list of defect locations and types which is sent to an output device for sorting.

### B. Radiological

Radiological sensing involves the use of x-ray radiation for IC inspection [4]. An x-ray inspection system typically has three parts: an x-ray source (tube), an image intensifier, and a fixture to hold and regulate the position of an inspection sample. Figure 5 shows an illustration of an x-ray inspection system.

### C. Acoustical

Acoustical sensing involves the use of sounds to identify defects in ICs. Scanning acoustic microscopy (SAM) [7] and surface acoustic waves (SAW) [8] are the two most widely used acoustical sensing modalities.

#### 1) Scanning Acoustic Microscopy (SAM)

Acoustic waves are utilized in SAM to provide visual images of changes in IC mechanical characteristics. SAM can measure properties such as density, thickness, roughness, stiffness, and attenuation. [7]. An illustration of a SAM system is shown in Figure 6. The main component of SAM is the transducer (usually a piezoelectric) which converts an electrical signal into an acoustic wave. The transducer is excited by an electrical tone burst, which transmits a packet of ultrasonic energy that is focused via a lens onto an IC under examination within a coupling medium. The IC reflects some of the ultrasound energy and sends it back to the transducer, where it forms an electrical signal, which is subsequently converted into an image.

#### 2) Surface Acoustic Waves (SAW)

Another acoustic sensing modality is surface acoustic waves (SAWs) [8, 9] which uses sound waves traveling along the surface of an IC. The amplitude of sound waves decreases upon penetrating deeper into the substrate. Most SAW inspection methods use non-contact laser excitation and sensing. As illustrated in Figure 7, a pulsed laser, a control and data acquisition unit, and a detector are the main components of a laser-based system. An IC generates ultrasonic waves by inducing heat with a pulsed laser as an excitation source. The transient out-of-plane displacement response is measured via a laser detector. The system finds flaws based on the responses received by measuring the response over the IC surface.

### D. Infrared Thermography

Infrared thermography (IRT) is a mechanism in which the heat that propagates in a solid is attenuated in space and is shifted in time [11]. It is a sensing modality which is commonly used for material evaluation [10]. The primary operating principle of IRT involves measuring the heat luminance from the surface in the electromagnetic spectrum region corresponding to the infrared (IR) wavelength and recording the temperature distribution of the surface. After a heating source is used to heat the IC under investigation, the defects in the IC modifies the thermal conductivity, resulting in a different temperature distribution. An illustration of an infrared thermography system is shown in Figure 8.

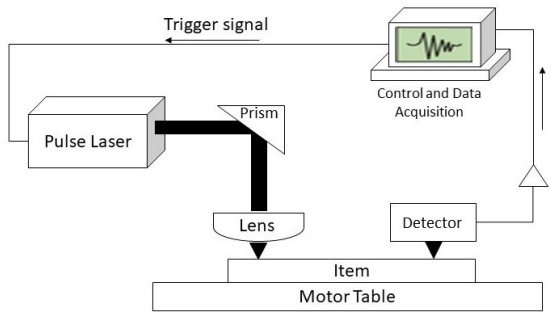

Fig. 7. Illustration of SAW inspection system

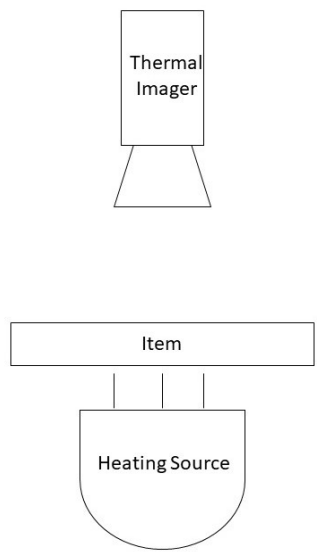

Fig. 8. Illustration of an IRT system

## IV. DETECTION METHODS FOR DIE ATTACHMENT AND WIRE BONDING DEFECTS

Die attachment and wire bonding are two crucial steps in IC manufacturing. Die attachment involves picking a die/chip from a wafer and placing it onto a substrate, a metal lead frame, or a printed circuit board (PCB). The way the die gets bonded sets its functionality. Solder alloys (eutectic and soft soldering), electrically conductive adhesives (ECA), and silver-based die attachment are the most widely used mechanisms [13].

Wire bonding is the process of attaching dies or chips to the terminals of a chip package or to a substrate directly. The bonding process establishes the metallic link between wires and substrate. The conventional packaging/assembly approach includes die attachment and wire bonding. However, other wire bonding technologies are also used where dies are soldered onto a substrate, a metal lead frame or a printed circuit board (PCB) using a die attachment material and for interconnection between die and substrate solder area, an array of bumps/balls is used [12]. This is known as flip chip bonding or flip chip ball grid array (FCBGA).

### A. Die Attachment Defects

Die attachment ensures mechanical reliability and thermal/electrical efficiency. It requires sufficient adhesion between the die and the substrate for maintaining resilience, high thermal conductivity, and low electrical resistivity. Some commonly encountered defects during die attachment are misplaced/misaligned die, excess/insufficient epoxy, and missing solder balls. Table I through Table V provide a listing of the papers reviewed.

*1) Division of Papers Based on Sensing Modality Used*
   *a) Automatic Optical Inspection (AOI)*

AOI has been extensively used to inspect surface defects related to die attachment [14] and solder joints [15-27]. In [14], three visual inspection stations behind a welding line site were used to collect data and analyze the differences in the characteristics of good and bad products by using image feature values towards the detection of skewed, drained, or offset die attachment defects. Many recent AOI inspection systems address detecting solder joints on PCBs [15-26] and high-density flexible IC substrates (FICS) [27] due to the difficulty in using a physical test-probe to access tiny and dense components on PCB and FICS for detecting defects. Furthermore, since the inspection operation for dense mounted boards are too complex and expensive [28], AOI systems for solder joints diagnosis is the modality widely used compared to the other sensing modalities.

   *b) X-ray*

X-ray is used to look for die attachment defects by its ability to penetrate the substrate material and detect hidden faults such as voids in die attachment [29], defective solder bumps/joints/balls in a ball greed array [30-37]. In [29], radiographic images were acquired from a semiconductor chip and then image processing methods were employed to automatically compute the amount of epoxy die attachment voids. In [30], a method for automatic detection and location of voids in die attachment solder joints was developed. In [31, 35], defective bumps/joints were detected. In [32], an inspection system was introduced to detect missing micro bumps from x-ray images. X-ray inspection techniques for different solder joint defects including cracks, voiding, head and pillow defects, missing solder, defects in bridges were studied in [33, 34, 36, 37].

*c) Scanning Acoustic Microscopy (SAM)*

SAM is extensively used for defect detection of the surface and internal microstructure of a flip chip. It is noted that recent SAM inspection approaches, e.g. [38-44], mainly focus on missing solder bumps/joints.

*d) Surface Acoustic Waves (SAW)*

SAW has been used to identify defects such as missing bumps and cracks in [45-49, 50], voids in micro solder bumps in [54, 55], underfills in [49], ball interconnections in flip chips and BGA in [51-53].

*e) Infrared Thermography*

Infrared thermography has been used to inspect flip chips to detect solder joint defects in [56, 58, 63], missing bumps in [57, 59, 61], cracks and voids in [60, 62].

*2) Division of Papers Based on Detection Method Used*

After acquiring image data via different sensing modalities, an inspection system often extracts features specific to a defect in order to perform its detection or classification. This is achieved via different image processing algorithms. In this section, recent works based on their detection methodology are reviewed.

*a) Image Processing*

In [14], the Otsu threshold edge detection method was used to obtain skewed, drained, and offset dies by using the horizontal projection pixel accumulation, Canny edge detection, and line Hough conversion methods. In [15], the Hue-Saturation-Intensity color space together with the Gabor transform were used to extract features for defect detection. In [16], image processing transformations of flipping, rotation, shifting, blurring, were used for preparation of image data. In [26], the pre-processing stage consisted of two steps of image segmentation for extracting the IC under inspection from a captured image by using a geometric feature vector.

Image data acquired via x-ray often use image processing algorithms such as filtering, energy normalization, thresholding, and image enhancement for defect detection as discussed in [29-37]. In [38-44], images of flip chips captured by SAM systems were first segmented based on the flip chip structure using different segmentation techniques.

*b) Conventional Machine Learning*

In [15], principal component analysis (PCA) was used for feature selection and then solder joints were classified using a support vector machine (SVM). In [17], a clustering-based classification method was presented in which active and semi-supervised learning were integrated for reducing the annotation workload. In [20], an IC solder joint inspection method was proposed based on an adaptive template scheme along with a weighting scheme to highlight defects in solder joint images. In [21], an IC solder joint inspection method was presented based on robust principal component analysis (RPCA) [21] and a Gaussian mixture model (GMM) [22]. SVM has been used in many works to perform defect classification [23]. In [24], a hierarchical multi-marked point model was developed for the task of detecting solder pastes scooping and scoop area estimation. In [25], a biologically inspired feature manifold framework was introduced for solder pastes defect detection and classification. In [26], a multi-layer perceptron neural network was designed for defect classification. In [27], curvature features extracted by an improved Weingarten mapping method were utilized to diagnose the quality of joints. Fuzzy support vector machine (F-SVM) and fuzzy C-means (FCM) algorithms were used for solder bump recognition in SAM in [38] and [43], respectively.

*c) Deep Learning*

A convolutional neural network was proposed in [16] to classify solder paste defects on PCBs. In [17], for solder joint localization, a deep ConvNet-based network named YOLO was utilized. In addition, transfer learning (VGG-16 model) and PCA were considered to achieve feature extraction and dimensionality reduction. In [18], a mask region-convolutional neural network (R-CNN) deep learning method was used to obtain image features for classification of solder joints. In [19], a cascaded convolutional neural network (CNN) was used for a similar purpose. In [31], a self-organizing map network was used for the recognition of missing bump defects from x-ray images. In [31], an ensemble based extreme learning machine was established to recognize defects. Different back propagation networks including Levenberg-Marquardt neural network (LM-BP) [44], general regression neural network (GRNN) [41], and radial basis function neural network (RBF) [42] have also been used for solder bump recognition based on SAM images.

*d) Other Methods*

The works done on post-processing analyses of SAW images include local temporal coherence analysis [45], wavelet analysis [46], spectral analysis [47] in the frequency domain, and correlation coefficient analysis [48-55] in the time domain. Infrared thermography thermal images and transient temperature response have also been used to analyze heat conduction for defect detection in [56-63]. In [58, 62], eddy current pulsed thermography (ECPT) was utilized to investigate thermal transfer through solder joints. In [63], k-means and in [61], a probabilistic neural network (PNN) were used to facilitate defect detection.

B. Wire Bonding Defects

Wire bonding defects refer to the type of defects that hinder electrical signals to be sent out of an IC such as broken wire, missing wire, etc. The papers on wire bonding defects mostly use the sensing modalities of AOI and x-ray. Table VI and Table VII provide a listing of the papers reviewed.

*1) Division of Papers Based on Sensing Modality Used*

*a) Automatic Optical Inspection (AOI)*

Due to recent advancements in AOI systems, they have become the preferred approach over the manual inspection approach. In [65], an AOI system was used to trace IC wires and to detect the bonding condition. In [14], along with detecting defects related to die attachment, wire bonding related defects such as wire breaks or leaks were detected using an AOI system (see Table 1). In [66], verifying the correctness of wire bonding positions was examined on a multi-layered IC wiring. In [67], an AOI system was used to detect broken, lost, shifted, shorted, and sagged wires.

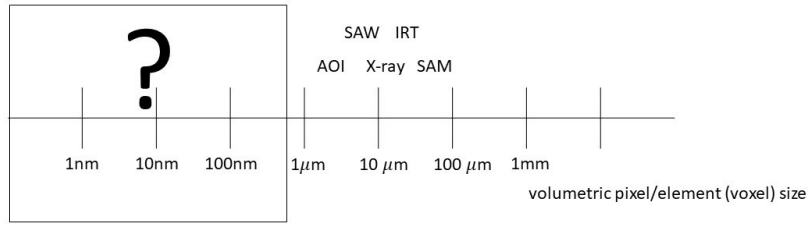

Fig. 9. Gap between IC size and spatial resolution of different sensing modalities

*b) X-ray*

Recently x-ray inspection was reported in [68-70] in which defects such as high/low loop, missing, sagged, and broken wires were identified.

*2) Division of Papers Based on Detection Method Used*

Similar to die attachment defect detection, wire bonding defect detection requires preprocessing, feature extraction and classification of acquired image data. The methods developed are mentioned in the following subsections.

*a) Image Processing*

To preprocess image data, the circle Hough transform (CHT) algorithm was used in [65] to locate bond balls. In [14], the Otsu threshold selection and connected-component algorithms were used to detect wire bonding defects. In [66], a method that integrated image processing algorithms (e.g., thresholding, binary labeling) was proposed together a wire bonding simulation to automatically inspect the correctness of wire bonding positions. In [67], a number of inspection algorithms based on image processing were covered along with a modified template matching algorithm to identify wire bonding defects. In [68], a transformation and threshold segmentation method was used to locate the region of interest (ROI) containing bond wires and a geometric feature extraction mechanism was utilized. In [69, 70], similar image processing algorithms were used for detecting wire bonding defects.

*b) Conventional Machine Learning*

Five popular machine learning methods of SVM, DT, RF, MLP, and KNN, were applied in [68] to identify chip image patches containing defective chips. A comparison was also carried out to select the best performed classifier for defect detection. The results were then verified by using a template matching algorithm.

*c) Deep Learning*

A CNN model together with a SVM classifier were used in [65] for the inspection of bonding balls. The experimental results obtained showed high classification outcomes. In [68], a template matching method together with a CNN model were used to validate the wire bonding inspection method reported.

Table VIII lists a summary or an overview of the reviewed papers in this section, providing a comparison of the existing methods in terms of their advantages and limitations.

## V. CHALLENGES AND POSSIBLE FUTURE RESEARCH DIRECTIONS

Quality assurance is critical in the IC manufacturing industry. Considering that ICs are becoming more dense, detecting defects is becoming more challenging. This section discusses challenges related to inspection systems together with future research directions for deploying them to detect die attachment and wire bonding defects in IC manufacturing.

*A. Challenges*

The major concern remains size reduction while retaining or improving performance at the same time. It is anticipated that the structural size of ICs become of the scale of a few nanometers [71]. The gap between the size of an IC and the spatial resolution of different sensing modalities is shown schematically in Figure 9. This indicates that a major challenge for detecting die attachment and wire bonding defects is a lack of spatial resolution to meet the sub $\mu$m scales.

Another challenge involves having an image acquisition component that can enables defect detection with high accuracies. Different sensing modalities require different hardware components. For example, AOI systems require proper lighting. As reported in [14], images associated with an AOI system are easily affected by bright light and improper shooting angle, and as a result their quality is impacted by such lighting noises. In other words, the practical setup of an image acquisition system remains a challenge that needs to be addressed further.

Feature selection is another challenge as it requires much experimentation to obtain effective features that allow separating defects from non-defects. The performance of an inspection system is highly dependent on the selected features [14, 23]. As the experimental results in [35] has shown, combining multiple features could enable improved detection outcomes.

The inspection of die attachment and wire bonding defects face additional hurdles for multilayered ICs due to the higher complexity in stackable dies and multilayer wire bonding. The defects in multilayered ICs are difficult to detect using image processing algorithms which are normally carried out in 2D since light from wires in higher layers cover wires in lower layers. In [66], a method was introduced to verify the correctness of the wire bonding positions on a multilayered wire IC.

Another challenge involves the speed or the fastness associated with a defect detection system so that it can work synchronously within the production line. Long inspection time are not deployable. Proper image acquisition, real-time image processing, and user-friendly graphical-user-interfaces are all needed for an in-line inspection system. Nearly all of the existing papers just focus on the detection accuracy without discussing these other critical deployment factors. Only in [66, 68, 70], the inspection speeds were reported.

*B. Possible Future Research Directions*

In this subsection, possible future research directions to improve the existing detection methods for die attachment and wire bonding defects are stated.

To address the aforementioned challenges regarding reduction of IC size and multilayered ICs, it is worth exploring the use of more than one sensing modality at the same time or combining two or more sensing modalities at the same time to increase the robustness of detection.

Since proper image feature selection plays a key role in obtaining high detection accuracies, a future research direction could be the design of more advanced deep learning feature selection methods, in particular features that take into consideration the limitation of image data for hard-to-access areas.

The real-time aspect of the entire inspection pipeline using modern fast processing engines such as GPUs (graphical processing units) and NPUs (neural processing units) is another future research direction that can be explored for the purpose of having a deployable inspection system in the manufacturing process.

VI. CONCLUSION

In this paper, a review or survey of the existing die attachment and wire bonding defect detection methods in integrated circuit manufacturing has been presented. Recent representative papers have been included in this survey. The papers have been organized in terms of the sensing modality used. Furthermore, the papers have been organized in terms of the image processing detection methods used: image processing, conventional machine learning, and deep learning. Tables have been put together providing an overview comparison among the papers reviewed. Challenges as well as possible future research directions for detecting die attachment and wire bonding defects have also been stated.

TABLE I. REPRESENTATIVE PAPERS OF AOI SYSTEMS FOR DETECTION OF DIE ATTACHMENT DEFECTS.

| Article # | Authors | Year | Type of Inspection | Detection Methodology | Performance | | | | |
|---|---|---|---|---|---|---|---|---|---|
| [14] | Chan et al. | 2018 | Attaching defects- skewed, drained, or offset and wire bonding defects- wire breaks or leaks | Image processing connected component method | Inspection type | Sample # | Precision | Recall | F-measure |
| | | | | | Defective substrate | 44 | 100% | 100% | 100% |
| | | | | | Die Attach Defects | 101 | 96% | 100% | 97.9% |
| | | | | | Wire Bonding Defects | 84 | 100% | 100% | 100% |
| [15] | Wu et al. | 2021 | Solder joints | Gabor features (feature extraction); principal component analysis (feature selection); SVM (classification) | Sample # | Training | | Testing | |
| | | | | | | 75 | | 30 | |
| | | | | | Feature extraction | Gabor features | | Recognition rate | |
| | | | | | | Real part | | 60% | |
| | | | | | | Imaginary part | | 80% | |
| | | | | | | Complex | | 62.9% | |
| | | | | | Classifier (SVM) | Solder joint type | | Correct Rate | |
| | | | | | | Good solder | | 100% | |
| | | | | | | Tombstone | | 75% | |
| | | | | | | Wrong component | | 100% | |
| | | | | | | No component | | 100% | |
| | | | | | | Pseudo solder | | 57.1% | |
| | | | | | | Component shifted | | 57.1% | |
| | | | | | | Total | | 80% | |
| [16] | Sezer and Altana | 2021 | Solder pastes defects | Image processing Optimization-based deep learning model-CNN | Sample # | Training | | Testing | |
| | | | | | | 582 | | 66 | |
| | | | | | Accuracy | 83% | | | |
| | | | | | Sensitivity | 67% | | | |
| | | | | | Specificity | 100% | | | |
| | | | | | Precision | 100% | | | |
| | | | | | Recall | 40% | | | |
| | | | | | F-measure | 80% | | | |
| [17] | Dai et al. | 2020 | Solder joint defects | Active and semi-supervised method (classification) Deep ConvNet-based method; YOLO algorithm (localization) | | Sample# | Precision | Recall | |
| | | | | | Dataset 1 | 5037 | 95.1% | 94.7% | |
| | | | | | Dataset 2 | 2543 | 87.6% | 94.6% | |
| [18] | Wu et al. | 2020 | Solder joint - incorrect | Mask Region-convolutional | Sample# | Training | | Testing | |
| | | | | | | 60 | | 28 | |

| Ref | Authors | Year | Target | Method | Metric | Value | | | |
|---|---|---|---|---|---|---|---|---|---|
| | | | component, proper component, component shift. Tombstone - no component. | neural network (R-CNN) | Accuracy | 100% | | | |
| | | | | | Mean of average precision (mAP) | 96.4% | | | |
| [19] | Cai et al. | 2018 | Solder joints | Cascade CNN | Sample# | Training 1108 | Testing 894 | | |
| | | | | | Recall | 99.8% | | | |
| | | | | | Precision | 100% | | | |
| | | | | | F-measure | 99.9% | | | |
| [20] | Ye et al. | 2018 | Solder joints | Adaptive template method | Sample# | 575 | | | |
| | | | | | Error rate | 0% | | | |
| | | | | | Omission rate | 0% | | | |
| [21] | Cai et al. | 2017 | Solder joints | Robust principle component analysis (RPCA) | Sample# | 574 | | | |
| | | | | | Error rate | 0.72% | | | |
| | | | | | Omission rate | 0% | | | |
| [22] | Cai et al. | 2016 | Solder joints | Gaussian mixture model (GMM) | Sample# | 575 | | | |
| | | | | | Training time | 583.566 ms | | | |
| | | | | | Inspecting time | 294.373 ms | | | |
| | | | | | Error rate | 1.04% | | | |
| | | | | | Omission rate | 0% | | | |
| [23] | Wu et al. | 2013 | | Color and template matching feature (feature extraction); Feature selection method based on information gain two-stage classifier-Bayesian and SVM | Sample# | 280 | | | |
| | | | | | Modeling time(s) | 0.13s | | | |
| | | | | | Correct rate | 100% | | | |
| [24] | Benedek et al. | 2013 | Solder Paste Scooping | Hierarchical multi-marked point process model | Sample# | 125 | | | |
| [25] | Jiang et al. | 2012 | Solder Paste Defects | Biologically inspired color feature (BICF); sub-manifold learning method; positive–negative discriminative analysis (PNDA) | Sample# | 40 (package 1005) | | | |
| | | | | | Number of training | 13 | 15 | 17 | 19 |
| | | | | | Recognition rate | 90.6% | 91.6% | 97% | 97.5% |
| [26] | Acciani et al. | 2007 | Solder joint defects | Geometric parameters (feature extraction); MLP neural network | Sample# | 656 | | | |
| | | | | | Recognition rate | Validation set no. 1 | 98.9% | | |
| | | | | | | Validation set no. 2 | 100% | | |
| | | | | | | Validation set no. 3 | 93.9% | | |
| [27] | Zeng et al. | 2021 | Solder joints | curvature and geometry features; pattern based discrete matching and classification method | Sample# | 40000 | | | |
| | | | | | Success rate | 98.3% | | | |

TABLE II. REPRESENTATIVE PAPERS OF X-RAY SYSTEMS FOR DETECTION OF DIE ATTACHMENT DEFECTS.

| Article # | Authors | Year | Type of Inspection | Detection Methodology | Performance | | | | |
|---|---|---|---|---|---|---|---|---|---|
| [29] | Kovac et al. | 2016 | Die attament defect - void | Image processing and statistical analysis | - | | | | |
| [30] | Amza | 2014 | Epoxy die attachment defect - void | Image processing Back tracking algorithm | Sample# | 49 | | | |
| | | | | | # of classes for segmentation | Performance rate | | | |
| | | | | | 2 | 78.3% | | | |
| | | | | | 3 | 92.3% | | | |
| | | | | | 4 | 86.4% | | | |
| [31] | Su et al. | 2019 | Solder bumps | Image processing Ensemble-ELM | Sample# for training | 4 (484 solder bumps) | | | |
| | | | | | Recognition rate | 97.9% | | | |
| [32] | Liao et al. | 2015 | Missing-bump defects | Image processing Self-organizing map (SOM) neural network | Sample# for training | 4 (454 solder bumps) | | | |
| | | | | | Recognition rate | 100% | | | |
| [33] | Lall et al. | 2014 | Solder joints | Image processing | - | | | | |
| [34] | Li et al. | 2011 | Solder bump bridging | Image processing | - | | | | |
| [35] | Teramoto et al. | 2007 | Solder joints | Image processing; Linear Discriminate Analysis (LDA) Artificial Neural Network (ANN) | Sample# | 1 (# of bumps 344) | | | |
| | | | | | Classifier | Correct rate | | | |
| | | | | | LDA | 99.7% | | | |
| | | | | | ANN | 99.7% | | | |
| [36] | Rooks et al. | 1995 | Ball joint defects | Image processing | Defect type | Joint# | False-alarm rate | Escape rate | Inconsistency rate |
| | | | | | Open/low solder | 3750 | 0.87% | 2.1% | 0.38% |
| | | | | | Open joints | 3750 | 0.081% | 0.0% | 0.029% |
| | | | | | Pad nonwets | 2500 | 0.37% | 0.0% | 0.18% |
| [37] | Roh et al. | 1999 | Solder joints | Image processing Learning vector quantization (LVQ) neural network and a lookup table (LUT) | Joint # | 68 | | | |
| | | | | | Success rate | 97% | | | |

TABLE III. REPRSENTATIVE PAPERS ON SAM SYSTEMS FOR DETECTION OF DIE ATTACHMENT DEFECTS.

| Article # | Authors | Year | Type of Inspection | Detection Methodology | Performance | | | | | | |
|---|---|---|---|---|---|---|---|---|---|---|---|
| [38] | Fan et al. | 2016 | Solder bumps | Fuzzy support vector machine (F-SVM) | Accuracy | 98% | | | | | |
| [39] | Su et al. | 2013 | Solder joints | Image Processing Backpropagation network | Sample chip | A | B | C | D | E | F |
| | | | | | Error rate | 5.99% | 1.26% | 0.95% | 5.68% | 3.47% | 0.63% |
| [40] | Tismer et al. | 2013 | Interconnect defects | Split spectrum analysis | - | - | | | | | |
| [41] | Wang et al. | 2019 | Solder defects | General regression neural network (GRNN) | Accuracy | 97.7% | | | | | |
| [42] | Liu et al. | 2018 | Missing micro bump defects | Radial basis function neural network (RBF) | Accuracy | 99% | | | | | |
| [43] | Lu et al. | 2018 | Solder bumps | Fuzzy C-means (FCM) | Accuracy | 94.3% | | | | | |
| [44] | Liu et al. | 2017 | Solder defects | Levenberg-Marquardt back-propagation network (LM-BP) | Flip chip# | Error rate | | | | | |
| | | | | | 1 | 1.58% | | | | | |
| | | | | | 2 | 4.73% | | | | | |
| | | | | | 3 | 4.10% | | | | | |
| [45] | Yang and Umme | 2009 | Solder joint/bump defects including missing, misaligned, open, and cracked solder joints/bumps | Local temporal coherence (LTC) analysis | - | - | | | | | |

TABLE IV. REPRESENTATIVE PAPERS OF SAW SYSTEMS FOR DETECTION OF DIE ATTACHMENT DEFECTS.

| Article # | Authors | Year | Type of Inspection | Detection Methodology | Performance |
|---|---|---|---|---|---|
| [46] | Yang et al. | 2006 | Solder bumps | Wavelet analysis | |
| [47] | Liu et al. | 2000 | Missing solder balls | Spectral analysis | |
| [48] | Gong et al. | 2013 | solder bump defects including crack and open | Modified correlation coefficient (MCC) | Validated by cross-sectioning method |
| [49] | Gong et al. | 2013 | Poor wetted solder bumps | Modified correlation coefficient (MCC) | Validated by cross-sectioning method |
| [50] | Ume et al. | 2011 | Solder bumps | Modified correlation coefficient (MCC) | Validated by comparison with electrical test and x-ray technique results |
| [51] | Reddy et al. | 2018 | Defects in 2nd level interconnects | Modified correlation coefficient (MCC) | Validated using cross-sectioning method |
| [52] | Reddy et al. | 2021 | Quality of solder ball interconnections | Modified correlation coefficient (MCC) | Validated with electrical testing, SEM, and dye and pry results |
| [53] | Reddy et al. | 2021 | Solder ball interconnection quality | Modified correlation coefficient (MCC) | - |
| [54] | Ume and Gong | 2013 | Solder joint voids | Modified correlation coefficient (MCC) | Validated using cross-sectioning method |
| [55] | Steen et al. | 2005 | Subsurface Defects- voids in the epoxy under-fill or solder balls | Modified correlation coefficient (MCC) | Validated using SAM images |

TABLE V. REPRESENTATIVE PAPERS OF INFRARED THERMOGRAPHY SYSTEMS FOR DETECTION OF DIE ATTACHMENT DEFECTS.

| Article # | Authors | Year | Type of Inspection | Detection Methodology | Performance |
|---|---|---|---|---|---|
| [56] | Chai et al. | 2003 | Solder joint defect | Active transient thermography | Validated by comparision of daisy chain resistance |
| [57] | Lu et al. | 2011 | Defects of solder joints - missing bumps | Active transient thermography | Validated using phase comparison between defect area and sound area |
| [58] | Zhou et al. | 2017 | Solder joint defects | Eddy current pulsed thermography (ECPT) | Validated by comparison of thermal image and transient temperature response |
| [59] | Lu et al. | 2014 | Solder bumps - missing bump | Analysis of transient response | Validated by comparison of temperature difference |
| [60] | Lu et al. | 2012 | Crack and void in solder bumps | Active transient thermography | Validated by comparison of thermal resistances of solder bumps |
| [61] | He et al. | 2017 | Missed solder balls and bumps | Principal component analysis (PCA) Probabilistic neural network (PNN) | Validated by comparison of expected and actual output |
| [62] | Zhou et al. | 2015 | Solder ball defects such as cracks, voids, etc. | Eddy current pulsed thermography (ECPT) | Validated by comparison of thermal image and transient temperature response |
| [63] | Lu et al. | 2018 | Mirco solder ball defects | K-means algorithm | Validated by comparison of expected and actual output |

TABLE VI. REPREENTATIVE PAPERS OF AOI SYSTEMS FOR DETECTION OF WIRE BONDING DEFECTS.

| Article # | Authors | Year | Type of Inspection | Detection Methodology | Performance | |
|---|---|---|---|---|---|---|
| [64] | Chan et al. | 2021 | Ball bonding condition (wire) | Circle hough transform algorithm (CHT) support vector machine (SVM) circle hough transform algorithm (CHT) Convolution neural network (CNN) Human judgement | Sample# | 494 |
| | | | | | Precision | 97.5% |
| | | | | | Recall | 99.4% |
| | | | | | F-measure | 98.5 |
| [65] | Long et al. | 2019 | Wire bonding joints | Image processing - PCA and SVM | Sample# | Training 500 / Testing 88 |
| | | | | | Accuracy | 97.3% |
| [66] | Perng et al. | 2010 | Wire bonding positions on a multi-layered wire IC | Image processing and wire bonding simulation | Mal-detection rate | 0 |
| | | | | | Lost detection rate | 0 |
| | | | | | Inspection speed | 0.0112 s per wire |
| [67] | Perng et al. | 2007 | Wire bonding defects including broken, lost, shifted, shorted, or sagged wires | Modified pattern matching method for localization and a set of algorithms for classification | - | |

TABLE VII. REPRESENTATIVE PAPERS ON X-RAY SYSTEMS FOR DETECTION OF WIRE BONDING DEFECTS.

| Article # | Authors | Year | Type of Inspection | Detection Methodology | Performance | |
|---|---|---|---|---|---|---|
| [68] | Chen et al. | 2021 | Wire bonding- normal bond wire; wire with high loop and low loop; sagged wire; broken wire; wire missing | Image processing Data driven-CNN | Sensitivity | 95% |
| | | | | | Accuracy | 93.6% |
| | | | | | Average time | 0.078s |
| [69] | Wang | 2002 | Wire bonding defects - bonding line missing and bonding line breakage | Image processing | Illumination level | Recognition rate |
| | | | | | 20 klx | 97.5% |
| | | | | | 40 klx | 98.7% |
| | | | | | 80 klx | 98.1% |
| | | | | | 90 klx | 99.7% |
| [70] | Tsukahara et al. | 1988 | Wire bonding defects - broken wires, too close wires, and incorrect wiring paths | Image processing | Inspection time | 28s (68-wire IC) |
| | | | | | Decoction accuracy | 99.7% |

TABLE VIII. SENSING MODALITIES USED FOR DETECTION OF DIE ATTACHMENT AND WIRE BONDING DEFECTS.

| Sensing Modality | Type of defect | Advantages | Limitations | Decision making method used |
|---|---|---|---|---|
| AOI | Die attachment, solder joints and wire bonding defects | • Low cost<br>• High data acquisition speed<br>• Suitable for detection of surface defects and flaws<br>• Noncontact and non-destructive | • Unable to detect inner defects | Machine vision techniques and machine learning (both conventional and deep learning) |
| X-Ray | Die attachment, solder bumps and wire bonding defects | • Used for both surface and inner inspection | • Traditional x-ray method can be destructive<br>• Micron level inspection has a low resolution<br>• Processing time is lengthy | Image processing and machine learning (both conventional and deep learning) |
| SAM | Solder bump defects | • Noncontact | • Micron and sub-micron level inspection has a low resolution<br>• Not suitable for in-line inspection<br>• Resolution vs. penetration depth: a trade-off<br>• Coupling medium is required. | Machine learning (both conventional and deep learning) |
| SAW | Solder bumps and solder ball interconnection defects | • Mostly noncontact | • Electronic packaging accessibility is critical.<br>• Sub-micron level inspection has a low resolution<br>• Less susceptible to deep defects<br>• Thin chips are more sensitive<br>• Requires reference data | Signal processing |
| Infrared Thermography | Solder ball defects | • Noncontact<br>• Suitable for both surface and inner inspection | • Thermal noise and overheating can be problematic<br>• Sub-micron level defect has weak signal difference<br>• Inspection thickness under the surface is limited | Machine learning (both conventional and deep learning) |